\documentclass[multicol,aps,prl,onecolumn,amssymb,showpacs,floatfix]{revtex4}
\usepackage{eucal,bm,graphicx}
\usepackage{gensymb,subfigure}
\usepackage{amsmath}
\usepackage{textcomp}

\begin{document}

  \date{ } \title{Inferring maps of forces inside cell membrane
  microdomains} \author{J.-B. Masson$^{a*}$, D. Casanova$^{b}$,
  S. T\"urkcan$^{b}$, G. Voisinne$^{a}$, M. R. Popoff$^{c}$,
  M. Vergassola$^{a}$, A. Alexandrou$^{b}$ \email[Corresponding
  author:
  ]{jbmasson@pasteur.fr}  \\
  \small{$^{a}$ Institut Pasteur, CNRS URA 2171, Unit In Silico
    Genetics, 75724 Paris Cedex 15, France} \\
  \small{$^{b}$Laboratoire d'Optique et Biosciences, Ecole Polytechnique, CNRS, INSERM, 91128 Palaiseau, France} \\
  \small{$^{c}$ Institut Pasteur, Bact\'eries ana\'erobies et Toxines,
    75724 Paris Cedex 15, France} \\}


\begin{abstract}
  Mapping of the forces on biomolecules in cell membranes has
  spurred the development of effective labels, e.g. organic fluorophores and
  nanoparticles, to track trajectories of single biomolecules.
  Standard methods use particular statistics, namely the
  mean square displacement, to analyze the underlying
  dynamics. Here, we introduce general inference methods to fully
  exploit information in the experimental trajectories,
  providing sharp estimates of the forces and the diffusion
  coefficients in membrane microdomains.  Rapid and reliable
  convergence of the inference scheme is demonstrated on trajectories
  generated numerically.  The
  method is then applied to infer forces and potentials acting on the
  receptor of the $\epsilon$-toxin labeled by lanthanide-ion
  nanoparticles. Our scheme is applicable to any labeled biomolecule
  and results show show its general relevance for membrane compartmentation.
\end{abstract}

\pacs{87.80.Nj, 02.50.Tt, 87.16.dp, 05.10.Gg}
\maketitle

The motion of proteins and lipids in cell membranes and its relation
to function have attracted considerable interest in recent years
\cite{Saxton08}. Motion is commonly followed by tracking of single
biomolecules labeled by an organic fluorophore or an inorganic
nanoparticle that allows detection via fluorescence, light scattering,
etc.  \cite{Saxton97}. Trajectories are usually analyzed by plotting
the mean-square displacement (MSD) as a function of time. Parameters
like diffusion coefficients and domain sizes are extracted by fitting
MSD curves to analytical behaviors expected for different modes of
motion, e.g.  free Brownian diffusion, directed, confined or anomalous
motion \cite{Saxton97}.

A major physical motivation to biomolecule tracking stems from the
actively debated origin of membrane compartmentation. Free diffusion
of membrane proteins in a sea of lipids was first postulated in the fluid
mosaic model \cite{Nicolson72}. Following experimental observations of
confinement, the lipid rafts \cite{Ikonen97} and the picket and fence
\cite{Kusumi05} models were proposed.  In the former, membrane
proteins are preferentially located in domains with different lipid
composition (lipid rafts).  In the latter, compartmentation is
ascribed to the combined action of the cytoskeleton and anchored
transmembrane proteins, forming fences and pickets,
respectively. Alternative models relying on more specific mechanisms
of protein-protein interactions have also been proposed
\cite{Salome03,Lang07}. Additional complexity arises from the fact
that different confinement mechanisms may coexist and depend on the
type of biomolecule \cite{Marguet06}.

The MSD-based approach has been used extensively. Alternative
observables related to first-passage times \cite{Klafter08} or radial
particle density distribution \cite{Verkman} have been proposed
recently.  More information on the dynamics is hidden in the full
trajectory of biomolecules, though.  Focusing on a single observable,
e.g. the second-order moment for MSD, has the virtue of simplicity yet
it wipes out information. In particular, it makes harder
discriminating among different models of motion and does not provide
systematic assessment of their validity.  A more general approach
based on inference methods \cite{Info_2} is taken by considering
the likelihoods of the models themselves.  A quantitative sense of
their validity is thus obtained, together with systematic estimates of
the parameters of the models and their uncertainties.

Our aim here is to present a general inference approach to obtain maps
of the forces and the potentials involved in the confined motion of
biomolecules in cell membranes. Inferences are shown to provide sharp
estimations of the local forces acting in the microdomains. We
specifically consider the case of the receptors of $\epsilon$-toxins
in the membrane of Madin-Darby canine kidney (MDCK) cells.

\medskip The $\epsilon$-toxin is responsible for lethal enterotoxemia
in livestock, due to the Gram-positive bacterium {\it Clostridium
  perfringens} (types B and D).  A relatively inactive peptidic
prototoxin is first synthesized and is then converted to a highly
potent mature protein by cleavage and removal of terminal amino
acids. The mature protein targets a specific receptor located
preferentially in detergent-resistent domains of MDCK cells
\cite{Popoff07}.  The protein acts by heptamerizing, which leads to the
formation of pores and the rapid modification of the membrane
permeability to ions, causing cell death without any entry of the
toxin into the cytosol \cite{Popoff01,Okabe02}.

To label the $\epsilon$-toxin, we used 30-50 nm amine-coated
lanthanide oxide nanoparticles (NPs)
$\text{Y}_{\text{0.6}}\text{Eu}_{\text{0.4}}\text{VO}_{\text{4}}$
(mean toxin:NP ratio, 1:1; see \cite{CasanovaJACS}). These
nanoparticles present several advantages: they are highly photostable
without emission intermittency, they are synthesized directly in water
and present extremely narrow emission, allowing efficient rejection of
cell fluorescence \cite{BeaurepaireNanoLett}. Their size is directly
determined from their luminosity \cite{CasanovaAPL}. Different
emission colors are obtained using different lanthanide ions
\cite{Buissette}.

We used a wide-field inverted microscope (Zeiss Axiovert 100) equipped
with a 63x, NA=1.4 oil-immersion objective and an EM-CCD (Roper
Scientific QuantEM:512SC). The Eu$^{3+}$ ions of the nanoparticles
were excited with the 465.8-nm line of an Ar$^+$-ion laser and their
emission was detected using a 617/8M filter (Chroma). MDCK cells were
grown to confluency on glass coverslips. They were then rinsed,
incubated with 0.04 nM of labeled $\epsilon$-toxin or prototoxin for
20 min, rinsed 3 times, and observed in Hanks buffer containing 1$\%$
fetal calf serum and 1$\%$ penicillin-streptomycin either at 20 or
30$\degree$C.

In all experiments ($\sim$400 cells), we observed several
nanoparticles bound to a specific receptor on the cell membrane. We
verified specificity of binding by pre-incubating the toxin for 1 h
with an $\epsilon$-toxin antibody that prevents binding to the
membrane (obtained as in \cite{Popoff01}) and verifying absence of
nanoparticles bound to the cells. Toxins were kept at concentrations
low enough to ensure that single toxins (and not oligomers) are
tracked. Trajectories similar to those of prototoxins, which do not
oligomerize, were indeed observed. The mean toxin:NP ratio 1:1
implies, assuming a Poisson distribution, that the fraction of NPs
bound to zero, one and two or more toxins are 37$\%$, 37$\%$ and
26$\%$, respectively. Nanoparticles without toxins do not bind to the
cells and are rinsed away. Given the size of the NPs, it is improbable
that more than one toxin is present on the same area of the NP surface
allowing simultaneous binding to more than one receptor. Furthermore,
the binding ability of a fraction of the toxins may be impaired by the
coupling to the NPs. We therefore estimate that the fraction of NPs
bound to more than one receptor is less than 10$\%$. We also labeled
$\epsilon$-toxins with the organic fluorophore Cy3 and observed again
similar trajectories. This implies that the nanoparticle label does
not modify the receptor motion, which is thus determined by the
receptor mass and the membrane characteristics (viscosity, forces,
etc.).  The receptor motion was studied during 150 to 300 s. Figure 1
shows a portion of the confined trajectory of a prototoxin bound to
its receptor.  We verified that, given the diffusion coefficient and
the domain size, we are not limited by the image acquisition time
(21.4 or 51.4ms) \cite{ KusumiBJ05,Salome06}. Relatively short
portions of trajectories were considered, so as to exclude possible
drifting of the membrane domain, the cell or the microscope setup.

\begin{figure}[tbp]
\begin{center}
\includegraphics[width=240pt]{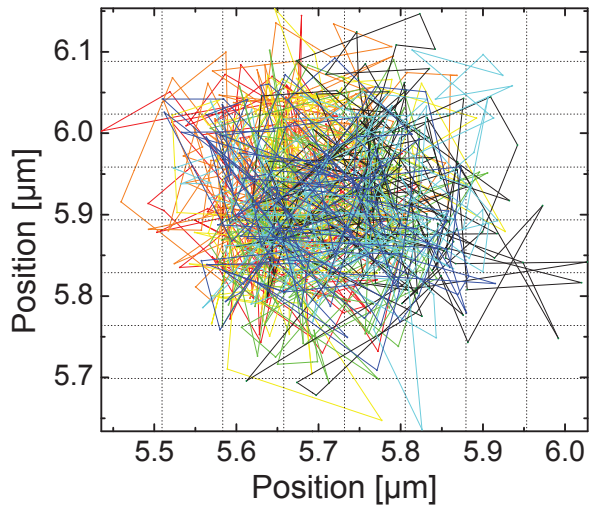}
\end{center}
\vskip -3.8mm
\caption{A 60-s trajectory of a
  $\text{Y}_{\text{0.6}}\text{Eu}_{\text{0.4}}\text{VO}_{\text{4}}$
  nanoparticle labeling $\epsilon$-prototoxin bound to its receptor on
  the membrane of an MDCK cell. The line color changes from red to
  blue (beginning/end of the trajectory). The motion is clearly
  confined.  See EPAPS document No. for the movie at
  real speed (scale bar: 1 $\mu$m).  Excitation intensity, 0.2
  kW/cm$^2$; integration time, 50 ms; readout time, 1.4 ms; $r$
  localization precision, 20 nm; temperature, 20$\degree$C.
  Inferred forces and potentials are shown in Fig. 2A. The dashed
  lines indicate the mesh squares used for the inference.}
\vskip -1.8mm
\label{figure_traj_1}
\end{figure}{\em }

\begin{figure}[tbp]
\begin{center}
\includegraphics[width=230pt]{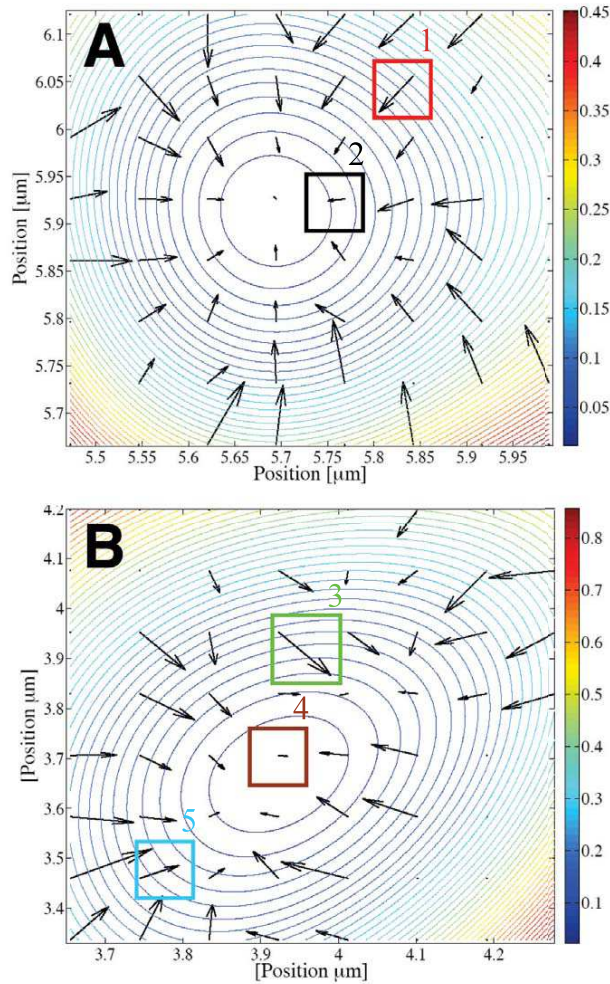}
\end{center}
\vskip -3.8mm
\caption{ Force and potential fields inferred inside two membrane
  microdomains. The length of the arrows is proportional to the
  magnitude of the force. The potential is plotted as level curves with the 
bar indicating the potential's amplitude on the isoline. The
  interpolation scheme described in the text was used,
  leading to an effective time step of 5 ms. The diffusion coefficent
  for \textbf{A} is $4.75\times 10^{-2} \mu m^{2}s^{-1}$ and for
  \textbf{B} is $8.15\times 10^{-2} \mu m^{2}s^{-1}$.
  \textit{Posterior} distributions for the five locations indicated by
  the squares are shown in Fig.~\ref{posteriori}.}  \vskip -1.8mm
\label{force_field1}
\end{figure}

\medskip Langevin equations for the position ${\bm r}(t)$ and the
velocity ${\bm v}(t)$ of a biomolecule subject to molecular diffusion
and to the force induced by a potential $V$ are\,:
\begin{equation}
\label{Langevin}
  \frac{d{\bm r}}{dt}={\bm v}\,;\quad m\frac{d{\bm v}}{dt}=
  -\gamma {\bm v}-\nabla V({\bm r}) +\sqrt{2D\gamma^{2}}{\bm \xi}\,.
\end{equation} 
Here, $m$ is the mass of the biomolecule, $\gamma$ and $D$ are the
friction and the diffusion coefficients inside the microdomain of the
membrane. The zero-average Gaussian noise $\xi(t)$ rapidly fluctuates
in time, accounting for the effect of thermal noise. Smoluchowski's
overdamped approximation \cite{schmo} to (\ref{Langevin}) is
sufficient for the motion of biomolecules.  Indeed, the typical time
for the relaxation of the velocity to local equilibrium is $\tau=
\frac{m}{\gamma}\simeq 10^{-16}s$ (since $m\simeq 10^{-22}\,kg$ and
$\gamma \simeq 10^{-6}\,kg/s$ \cite{Sheetz00}). Hence, the velocity is
slaved to its local forcing and (\ref{Langevin}) reduces to
\begin{equation}
\label{overdamped}
\frac{ d{\bm r}}{dt}=-\frac{\nabla V({\bm r})}{\gamma} +\sqrt{2D}{\bm \xi}.
\end{equation} 
The Fokker-Planck equation \cite{Risken} associated to
(\ref{overdamped}) reads
\begin{equation}
\label{FP}
\partial_t P= -\frac{1}{\gamma}{\bm \nabla}\cdot\left({\bm F}\,P\right) + 
D \Delta P\,,
\end{equation}
where the force ${\bm F}\equiv-{\bm \nabla}V$. Kolmogorov equation
(\ref{FP}) governs the transition probability $P({\bm r},t |{\bm
  r}_0,t_0)$ to get to the space-time point (${\bm r},t)$ conditional
to the initial space-time position $({\bm r}_0,t_0)$ of the
biomolecule.  It follows from (\ref{FP}) that the probability $P$ can
be expressed as a path integral \cite{Kleinert} over all
paths ${\bm r}(s)$ connecting ${\bm r}_0$ to ${\bm r}$\,:
\begin{equation}
\label{path_int}
P({\bm r},t |{\bm r}_0,t_0)\propto\int {\cal D}{\bm r}(s)\,
e^{-\int\,ds\, Q({\bm r}(s))}\,.
\end{equation} 
The term $Q({\bm r}(s))\equiv\left(d{\bm r}(s)/ds-{\bm F}({\bm
    r}(s))/\gamma\right)^{2}/4 D$ is the quadratic Gaussian weight
governing the probability of displacements over an infinitesimal time
interval.

In practice, space is discretized in a fine regular mesh of $n^{2}$
squares, as shown in Fig.~1 ($n=8$). The size of the mesh is taken
small enough for the forces to be smooth on that scale. At the lowest
order, forces are approximated by a constant value within each mesh
square and we shall show later that higher-order variations are indeed
negligible. The integral appearing at the exponential in
(\ref{path_int}) is approximated by the corresponding discrete Riemann
sum (see (\ref{sum_like})). Mesh squares ${\cal S}_{i,j}$ are indexed
by the pair $\left(i,j\right)$ (with $i,j=1,\ldots,n$) and the force
acting in ${\cal S}_{i,j}$ is denoted ${\bm F}_{i,j}$.  Our goal is to
estimate the $2n^{2}+1$ unknowns $U=\left\{D,\left\{{\bm
      F}_{i,j}\right\}\right\}$, i.e. the forces and the diffusivity
within a membrane subdomain, governing the trajectories of the labeled
biomolecule. Variations in $D$ can be handled similarly (see below).

Inferences methods (see, e.g., \cite{Info_2}) generally feature two
steps: a) the derivation of the \textit{posterior} probability
distribution of the unknown parameters of the model given the
experimental observations; b) sampling from the \textit{posterior}
distribution to estimate the parameters.  Specifically, it follows
from Bayes rule that the \textit{posterior} probability distribution
$P\left(U|T\right)$ of the set of unknown parameters $U$ given an
observed trajectory $T$ reads
\begin{equation}
\label{posterior}
P\left(U | T\right)=\frac{P\left(T | U\right)\times P_{0}\left(U\right)}
{P\left(T\right)}\,,
\end{equation} 
where $P\left(T | U\right)$ is the likelihood of a trajectory given
the parameters $U$ and $P\left(T\right)$ is a normalizing
constant. $P_{0}(U)$ is the \textit{prior} probability, which we take
constant. As for the sampling part (b), we used Monte Carlo methods to
compute the average over the \textit{posterior} distributions. The
latter are generally well-peaked and maximum values provide then good
estimates of the average values.

An asset of our specific problem is that the diffusivity $D$ is
the only global parameter whilst the $n^{2}$ forces ${\bm
  F}_{i,j}$ appear in the likelihood additively at the exponential. It
follows that the contributions of the various squares of the mesh
factorize as $P\left(U | T\right) = \prod_{i,j=1}^{n} P\left({\bm
    F}_{i,j},D|T\right)$. The contribution of each mesh square reads
\begin{equation}
\label{sum_like}
P\left({\bm F}_{i,j},D|T\right) \propto \prod_{\mu : 
{\bm r}_{\mu}\in {\cal S}_{i,j}}
\frac{\exp\left[-\frac{\left({\bm r}_{\mu+1}
        -{\bm r}_{\mu}-{\bm F}_{i,j}\Delta t/\gamma \right)^{2}}{4D\Delta t}\right]}{4\pi D\Delta t}\,.
 \end{equation}
 Here, $\mu$ indexes the various time steps (discretized by $\Delta
 t$) and the product is restricted to those times when the biomolecule
 is detected within the mesh square ${\cal S}_{i,j}$. Note that
 discretization introduces {\it a fortiori} an ambiguity when the
 biomolecule crosses the lines of the mesh and moves to a new
 square. The choice made in (\ref{sum_like}) is to simply use indices
 of the starting square. Corrections will be shown shortly to be
 negligible.

\begin{figure}[tbp]
\begin{center}
\includegraphics[width=230pt]{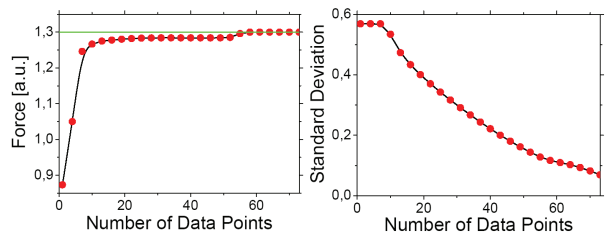}
\end{center}
\vskip -2.8mm
\caption{Typical evolution of the inferred value of a local force
  (left) and its standard deviation (right) with the
  number of points used to infer them. Note the rapid convergence
  to the real value of the force (indicated by the
  solid horizontal line). }
\label{figure_convergence_1}
\end{figure}
 
The crucial element ensuring well-peaked {\it posterior} distributions
and sharp inferences is that the trajectories of the biomolecules are
well confined to subdomains. It follows that the various squares of
the mesh are crossed multiple times, permitting the acquisition of a
massive amount of information. Even for those squares where the
largest forces are measured, i.e. the residence time is the shortest,
the amount of data is sufficient to permit sharp inferences. Note also
that {\it posterior} distributions for the forces are Gaussian, as
seen directly in (\ref{sum_like}).

\medskip To have a quantitative sense of the quality of the inference
scheme, we numerically generated ensembles of trajectories with the
same force fields and diffusion coefficients as obtained from the
experimental data.  \textit{Posterior} distributions were found to be
sharply peaked at the values used to generate the trajectories.
Typical evolutions of inferred values {\it vs} the number of
points used for the inference are shown in
Fig.~\ref{figure_convergence_1}. Convergence is manifestly rapid
and the standard deviation brackets the real value even for
few data points, providing a sensible estimate of the error
bars. Predictions by our method were found to be more precise and to
require less data points than those based on a single statistic,
e.g. MSD or radial particle density distribution. In summary, simulations provide strong
support to the validity of the inference method.

\medskip To visualize the results, it is convenient to plot the
potentials $V$ as in Fig.~\ref{force_field1}. To that purpose, the
potential is written as a polynomial of order $C$ ($C=4$ in
Fig.~\ref{force_field1})\,: $V\left({\bm r}\right)=
\sum^{C}_{j=0}\sum^{j}_{i=0}\alpha_{ij}x^{i}y^{j-i}$.  The constants
$\alpha_{ij}$ are fitted to the experimental force fields, minimizing
the squared error by standard simplex methods.  Potentials that we
find are incompatible with a cytoskeleton fence-type model, where a
steep wall-like potential is expected. This type of domain, however,
may still influence the receptor trajectories on time scales below our
resolution. Variations within microdomains for the diffusivities were
found to be small, i.e. about $4\%$ {\it vs} $65\%$ for the
forces.

\begin{figure}[tbp]
\begin{center}
\includegraphics[width=230pt]{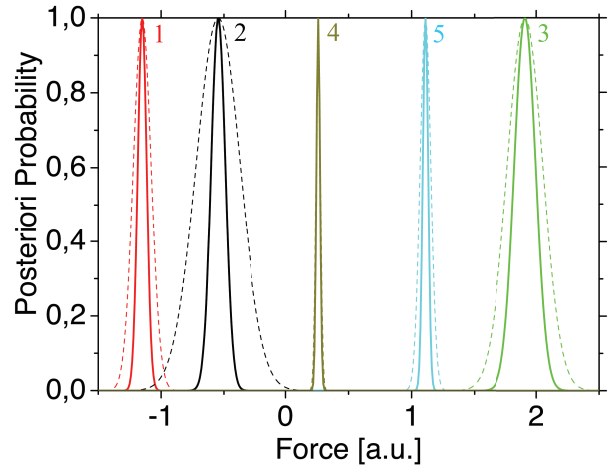}
\end{center}
\vskip -3.8mm
\caption{{\it Posterior} probability distributions of the forces at
  the locations indicated with the squares in Fig.~\ref{force_field1}. The
  curves for horizontal (vertical) components of the force are 2,4,5 
  (1,3). Solid curves are obtained by
  the interpolation scheme described
  in the text. Note that average values are extremely close to those
  obtained without interpolation (dashed curves) yet the variance is
  reduced.}
\vskip -2.mm
\label{posteriori}
\end{figure}

Discretization errors were controlled by the following method. Given
two acquisitions $({\bm x}_1,t_1)$ and $({\bm x}_2,t_2=t_1+\Delta t)$,
we interpolate their transition probability by summing over all possible
positions ${\bm x}'$ at the intermediate time $t'=t_1+\Delta t/2$,
i.e. $P\left(x_{2},t_{2} | x_{1},t_{1} \right) = \int dx'
P\left(x_{2},t_{2} | x',t'\right) P\left(x',t' |
  x_{1},t_{1}\right)$. The process can be further refined by
introducing additional intermediate points. The effect of the
interpolation mostly amounts to a reduction of the error bars, without
any major shift in the estimates of the forces, as can be seen in
Fig.~\ref{posteriori}.

In conclusion, we have developed an inference approach that fully
exploits information hidden in labeled biomolecule trajectories.  The
technique is generally applicable to any type of biomolecule and
trajectory, including intermittent trajectories like those obtained
with blinking quantum dots, and for forces and domains that change in
time. We have explicitly demonstrated the value of the method by
mapping the forces and the potentials involved in the confined motion
of the $\epsilon$-toxin receptor in the membrane of MDCK
cells. Results obtained here indicate that the method, especially in
combination with data on cytoskeleton destruction and cholesterol
depletion, is poised to shed light onto the controversial mechanisms
of membrane compartmentation.

{\bf Acknowledgments} We are grateful to G. Mialon, T. Gacoin,
J.-P. Boilot for the amine-coated nanoparticles, to C. Bouzigues for
helpful discussions, to the Region Ile-de-France Nanosciences
Competence Center, the Fonds National de la Science (ACI DRAB), the ANR PNANO (Grant 062 03) and
the Delegation Generale de l'Armement (D. C.) for financial support.

\bibliographystyle{apsrev}
\bibliography{masson}
$^{*}$ Corresponding author: jbmasson@pasteur.fr

\end{document}